\documentclass[%
prl, reprint,
superscriptaddress,
 amsmath,amssymb,
 aps,
]{revtex4-1}

\usepackage{graphicx}
\usepackage{dcolumn}
\usepackage{bm}
\usepackage{color} 
\usepackage[normalem]{ulem}
\usepackage{hyperref}
\usepackage[english]{babel}

\begin{document}

\title{Cross-Kerr nonlinearity for phonon counting}
\author{Shiqian Ding}
\altaffiliation{Present address: JILA, National Institute of Standards and Technology and University of Colorado, and Department of Physics, University of Colorado, Boulder, CO 80309, USA}
\author{Gleb Maslennikov}
\author{Roland Habl{\"u}tzel}
\affiliation{Centre for Quantum Technologies, National University of Singapore, 3 Science Dr 2, 117543, Singapore}
\author{Dzmitry Matsukevich}
\affiliation{Centre for Quantum Technologies, National University of Singapore, 3 Science Dr 2, 117543, Singapore}
\affiliation{Department of Physics, National University of Singapore, 2 Science Dr 3, 117551, Singapore}

\date{\today}

\begin{abstract}
State measurement of a quantum harmonic oscillator is essential in quantum optics and quantum information processing. In a system of trapped ions, we experimentally demonstrate the projective measurement of the state of the ions' motional mode via an effective cross-Kerr coupling to another motional mode.  This coupling is induced by the intrinsic nonlinearity of the Coulomb interaction between the ions. We spectroscopically resolve the frequency shift of the motional sideband of the first mode due to presence of single phonons in the second mode and use it to reconstruct the phonon number distribution of the second mode.
\end{abstract}

\pacs{37.10.Ty, 05.45.Xt, 03.67.Lx}
\maketitle
The quantum harmonic oscillator is one of the foundational models in physics which describes, among many other systems,
the mode of the electromagnetic field and the motion of trapped particles. A rich toolbox of methods exists to 
characterize its quantum state. In optics such methods include homodyning 
\cite{lvovsky_2001}, photon counting~\cite{Mandell_book} and photon number resolving detection 
\cite{NIST_resolved_2003,Kardynal_resolved_2008}. In a trapped-ion system the motion of ions is usually probed 
by coupling it to the ion's internal state via a motional sideband transition, which enables reconstruction of phonon number distribution~\cite{1996PRLNonclassicalstate,Leibfried_tomo_1996,2014PRLDuanboson,2014natphyKiwan,um_phonon_2016} 
or measurement of the 
parity of the motional state~\cite{ourpaper_parametricoscillator}. However most of the methods to determine the motional state are destructive 
in nature and the state of the oscillator after measurement does not correspond to its outcome.  

An ideal projective measurement should leave the quantum system immediately after measurement in the state defined by the measurement outcome. 
Such measurements have been performed by utilizing the nonlinear dispersive interaction between the oscillator and another quantum system such as Rydberg atoms~\cite{Haroche_2002,Haroche_2007Nature}, superconducting circuits~\cite{Schoelkopf_2014,heeres_cqed_cavity_2015} or electron motion in a Penning trap~\cite{gabrielse_1999}. 
An example of such an interaction in the context of a projective measurement of the Fock state~\cite{milburn_1983,yamamoto_1985,munro_2005} is the cross-Kerr nonlinear coupling between two different quantum oscillators. This coupling is described by the Hamiltonian $\hat{H}_{kerr} \sim \chi \hat{n}_a \hat{n}_b$, where $\hat{n}_a = a^{\dagger} a$, $\hat{n}_b = b^{\dagger} b$ are the number operators and $a$ ($a^{\dagger}$) and $b$ ($b^{\dagger}$) are the annihilation (creation) operators for the oscillators. 
Such an interaction enables the implementation of quantum gates~\cite{yamamoto_1995}, 
entanglement distillation~\cite{duan_2000} and the preparation of nonclassical states~\cite{paternostro_cat_2003}.

However, several difficulties arise in the practical implementation of the cross-Kerr nonlinearity. 
In optics, the interaction between the photons is usually weak~\cite{venkataraman_2013} and difficult to control. 
In addition, locality and causality arguments preclude large conditional phase shifts for traveling 
single photon wave packets ~\cite{shapiro_kerr_2006,shapiro_kerr_2007,gea_banacloche_kerr_2010,fan_kerr_breakdown_2013}.
One way of overcoming these limitations was recently demonstrated by mapping one of the photons to an atomic 
excitation ~\cite{vuletic_2016,tiarkse_pi_shift_2016,liu_cross_phase_2016}. Other similar schemes have also been  proposed~\cite{brod_2016}.  

In our approach we simulate the cross-Kerr interaction in quantum optics in a system of trapped ions. The anharmonicity of the Coulomb interaction induces a strong nonlinear interaction between the motional modes~\cite{james_complete_2003}. In the dispersive regime, where no energy exchange between the modes is possible, this coupling manifests itself as a shift of the frequency of the motional mode that is proportional to the number of phonons in another motional mode and can be described by an effective cross-Kerr interaction~\cite{Nie_2009}. The motional modes of the ions are localized and therefore not subject to the no-go theorems mentioned above.

A small shift (about 20 Hz/phonon) of this origin was first observed
in \cite{Roos_shift_2008} using a Ramsey type experiment. 
In this Letter, we demonstrate an order of magnitude larger conditional shift on the order of 300 Hz/phonon. Such a large shift together with the
long coherence time of the motional mode
allows us to spectroscopically resolve distinct peaks in the sideband spectrum of the motional
mode for different Fock states and perform  projective measurements of another motional mode in the Fock basis.

We trap three $^{171}$Yb$^+$ ions in a standard linear rf-Paul trap~\cite{ourpaper_parametricoscillator,microwave} with the single ion secular frequencies $(\omega_x, \omega_y, \omega_z) = 2\pi \times (1042, 979, 587)$ kHz. 
The axial trapping frequency $\omega_z$ is fixed and the radial trapping frequencies $\omega_x, \omega_y$ can be fine tuned by adjusting the offset voltages 
applied to the trap electrodes. 
Two of the ions are optically pumped to a metastable long-lived $^2F_{7/2}$ state and remain dark
throughout the experiment. 
We employ standard optical pumping and resonance fluorescence techniques~\cite{2007PRA} to initialize and detect the internal  
$|^2S_{1/2}, F=1, m_F=0\rangle \equiv |{\uparrow} \rangle$  and 
$|^2S_{1/2}, F=0, m_F=0\rangle \equiv |{\downarrow} \rangle$ states of the remaining bright ion.

\begin{figure}[tb]
\centering
\includegraphics[width= \columnwidth]{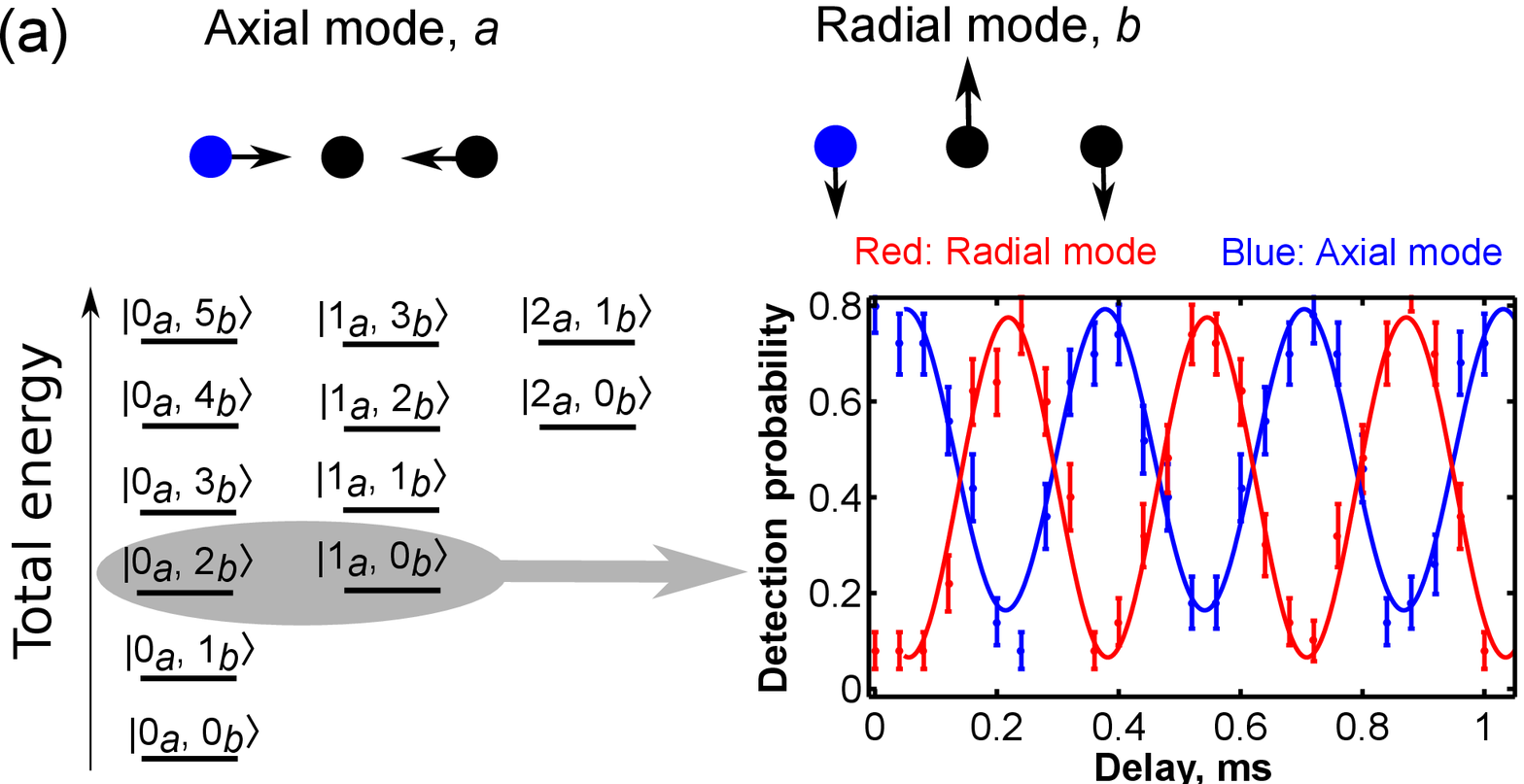}

\vspace{0.5cm}
\includegraphics[width=1.0 \columnwidth]{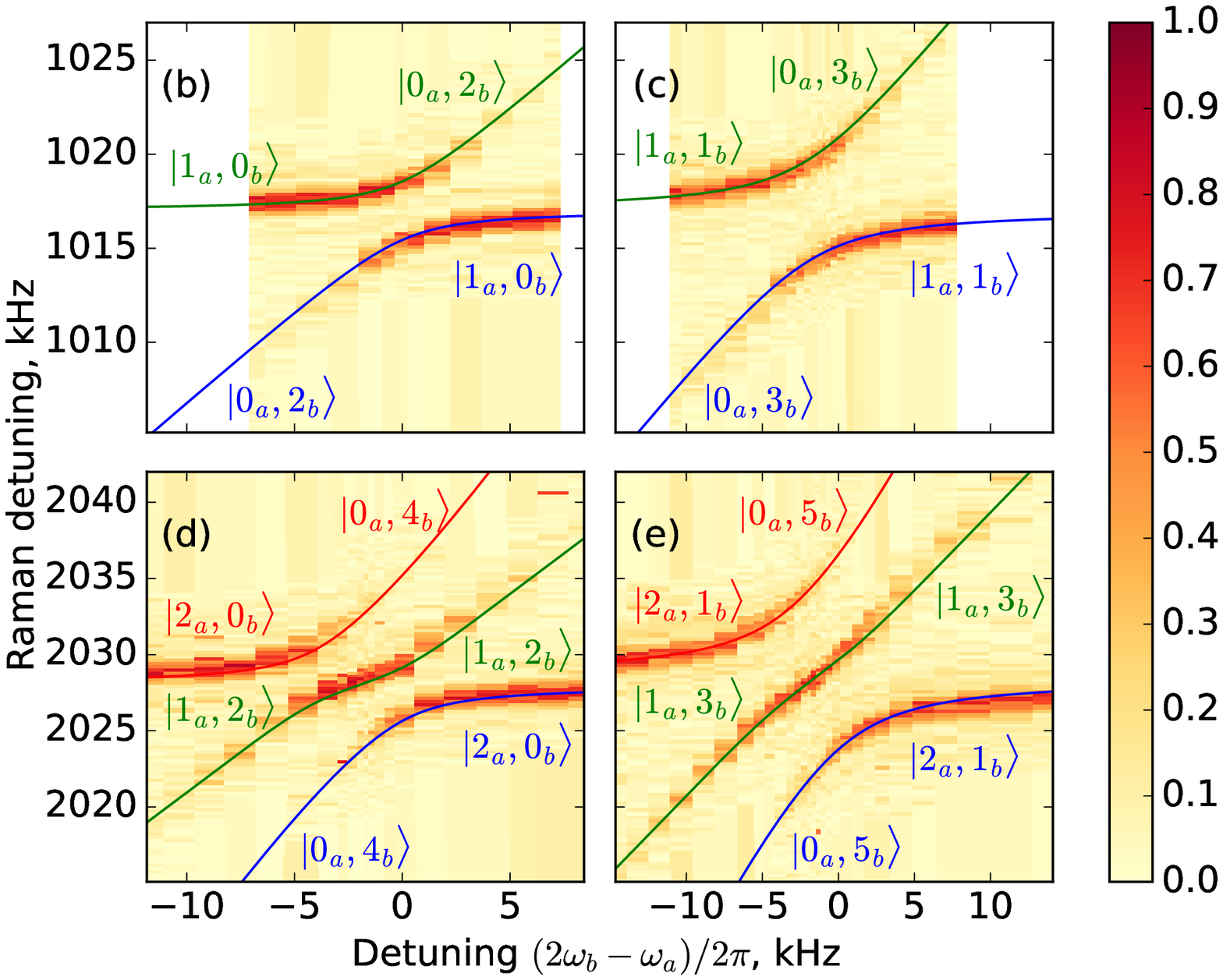}
\caption{\label{fig:anticrossing}
(a) Energy diagram of the ``bare'' motional eigenstates at resonance condition Eq.(\ref{eq:resonance}). The plot on the right  shows the coherent energy exchange between the axial and radial modes after we initialize the system in the state $|1_a, 0_b\rangle$. The blue (red) dots are the probability for the system to be in the  state $|1_a, 0_b\rangle$ ($|0_a, 2_b\rangle$) and the blue (red) line is the corresponding fit.
(b-e) Probability to find the bright ion in the state $|{\uparrow}\rangle$ as a function of Raman detuning and two-mode detuning $\delta$. Raman detuning is scanned around the first order (b,c) and the second order (d,e) motional blue sideband of the axial mode. 
The duration of Raman pulse is chosen to match the $\pi$ pulse of the corresponding blue sideband at large $\delta$.
Solid lines show the eigenen-energies of the Hamiltonian Eq.(\ref{eq:hamiltonian}).
}
\end{figure}

Three Raman beams are used to address all nine motional modes along all three principal axes of the trap
for sideband cooling, motional state preparation and detection. 
The Raman beams are produced by a frequency-doubled mode-locked Ti:Sapphire laser~\cite{Hayes_2010} with a repetition rate 
of 76.20 MHz and total power of around 250 mW at wavelength 374 nm. 
To ensure that all the motional modes can be addressed by the Raman lasers, 
the bright ion is always positioned at the edge of the crystal (see Fig.\ref{fig:anticrossing}(a)).
To achieve this, around $10\%$ of the collected photons is sent to an EM-CCD camera 
for constant monitoring of the position of the bright ion. If the ion jumps to the center of the crystal, 
for example, due to collisions with background gas, 
we melt and re-crystallize the ions by briefly interrupting the RF signal sent to the trap for a few microseconds 
to return the bright ion to the side of the crystal~\cite{shiqianThesis}.

Due to anharmonicity of the Coulomb interaction between the ions, the axial ``breathing" 
mode with eigenvector $e_a = (1, 0, -1) / \sqrt{2}$ and frequency $\omega_a=\sqrt{3}\omega_z$ is coupled 
to the radial ``zigzag" mode with eigenvector $e_b = (1, -2, 1) / \sqrt{6}$ and frequency $\omega_b=\sqrt{\omega_x^2-12\omega_z^2/5}$ (see Fig.\ref{fig:anticrossing}(a))~\cite{james_complete_2003}. 
Near the resonance condition 
\begin{equation}
\omega_a = 2 \omega_b,
\label{eq:resonance}
\end{equation}
by applying the rotating wave approximation and neglecting other off-resonant modes,  
we obtain the Hamiltonian that describes this coupling \cite{james_complete_2003,ourpaper_parametricoscillator}
\begin{equation}
H = \hbar \omega_a a^{\dagger} a + \hbar \omega_b b^{\dagger} b + \hbar \xi (a^{\dagger} b^2 + a b^{\dagger2}),
\label{eq:hamiltonian}
\end{equation}
where $a$ ($a^\dagger$) and $b$ ($b^\dagger$) are the annihilation (creation) operators for the axial and radial motional modes respectively,
$\xi = 9 \omega_z^2 \sqrt{\hbar /m \omega_a \omega_b^2} / 10 x_0 $ is the coupling strength
between the modes, $x_0 = (5 e^2 / 16 \pi \epsilon_0 m \omega_z^2)^{1/3}$ is the equilibrium distance between neighboring ions, and $m$ 
is the mass of one ion. 
Without the coupling term $H_c=\hbar \xi (a^{\dagger} b^2 + a b^{\dagger 2})$, the system consists of two independent harmonic oscillators and its
spectrum comprises a series of degenerate manifolds of states. The corresponding energy level diagram is plotted in Fig.\ref{fig:anticrossing}(a).
These ``bare" energy eigenstates are coupled by the term $H_c$, which leads to coherent energy exchange between motional modes  when the resonance condition  
Eq.(\ref{eq:resonance}) is satisfied, and the avoided crossing of the eigenenergies near the resonance~\cite{ourpaper_parametricoscillator}.

To verify this, both motional modes are initialized to ground state with detuning
$\delta = 2 \omega_b - \omega_a  = 2 \pi \times 88$~kHz. 
After adding one phonon to the axial mode by applying a $\pi$ pulse on the motional blue sideband transition, 
we adjust the voltages applied to the electrodes to bring the detuning $\delta$ to zero in around 25 $\mu s$, 
and let the system evolve for time $\tau$. 
We then return the detuning $\delta$ back to its initial value and detect motional excitation by mapping it to the ion internal state.
The results of our measurements are shown in the Fig.\ref{fig:anticrossing}(a). 
We observe oscillations between the single phonon state
in the axial mode $|1_a, 0_b\rangle$ and the two phonon state in the radial mode $|0_a, 2_b\rangle$,
with frequency $3.06(3)$kHz in good agreement with theoretical prediction
$2\langle 0_a, 2_b|H_c|1_a, 0_b\rangle/2\pi= 2 \sqrt{2} \xi / 2 \pi = 3.11$kHz.

To observe the avoided crossing, we start with both motional modes in the ground state and then measure the motional blue sideband of the axial mode in the vicinity of the resonance condition Eq.(\ref{eq:resonance}). The coupling $H_c$ mixes the axial and radial modes, such that all the eigenstates of the Hamiltonian Eq.(\ref{eq:hamiltonian}) have non-zero components along the axial direction and manifest themselves in the axial sideband scan near $\delta = 0$. Fig.\ref{fig:anticrossing}(b) shows the avoided crossing between the $|1_a,0_b\rangle$ and $|0_a,2_b\rangle$ eigenstates of the ``bare" Hamiltonian when the two mode detuning $\delta$ is scanned across the resonance. The avoided crossing between the eigenstates $|1_a,1_b\rangle$ and $|0_a,3_b\rangle$ shown in Fig.~\ref{fig:anticrossing}(c) is measured by preparing the motional modes in the initial state $|0_a,1_b\rangle$  before scanning the axial motional sideband. As expected, we observe a factor of $\sqrt[]{3}$ larger splitting.
The sideband spectrum in the vicinity of the second order motional sideband reveals avoided crossings 
between manifolds of three states ($|0_a,4_b\rangle$, $|1_a,2_b\rangle$, $|2_a,0_b\rangle$)
and ($|0_a,5_b\rangle$, $|1_a,3_b\rangle$, $|2_a,1_b\rangle$) as shown in  Fig.\ref{fig:anticrossing}(d) and (e) respectively. 
Similarly, the splitting is larger in the latter case.
\begin{figure}[tb]
\centering
\includegraphics[width=\columnwidth]{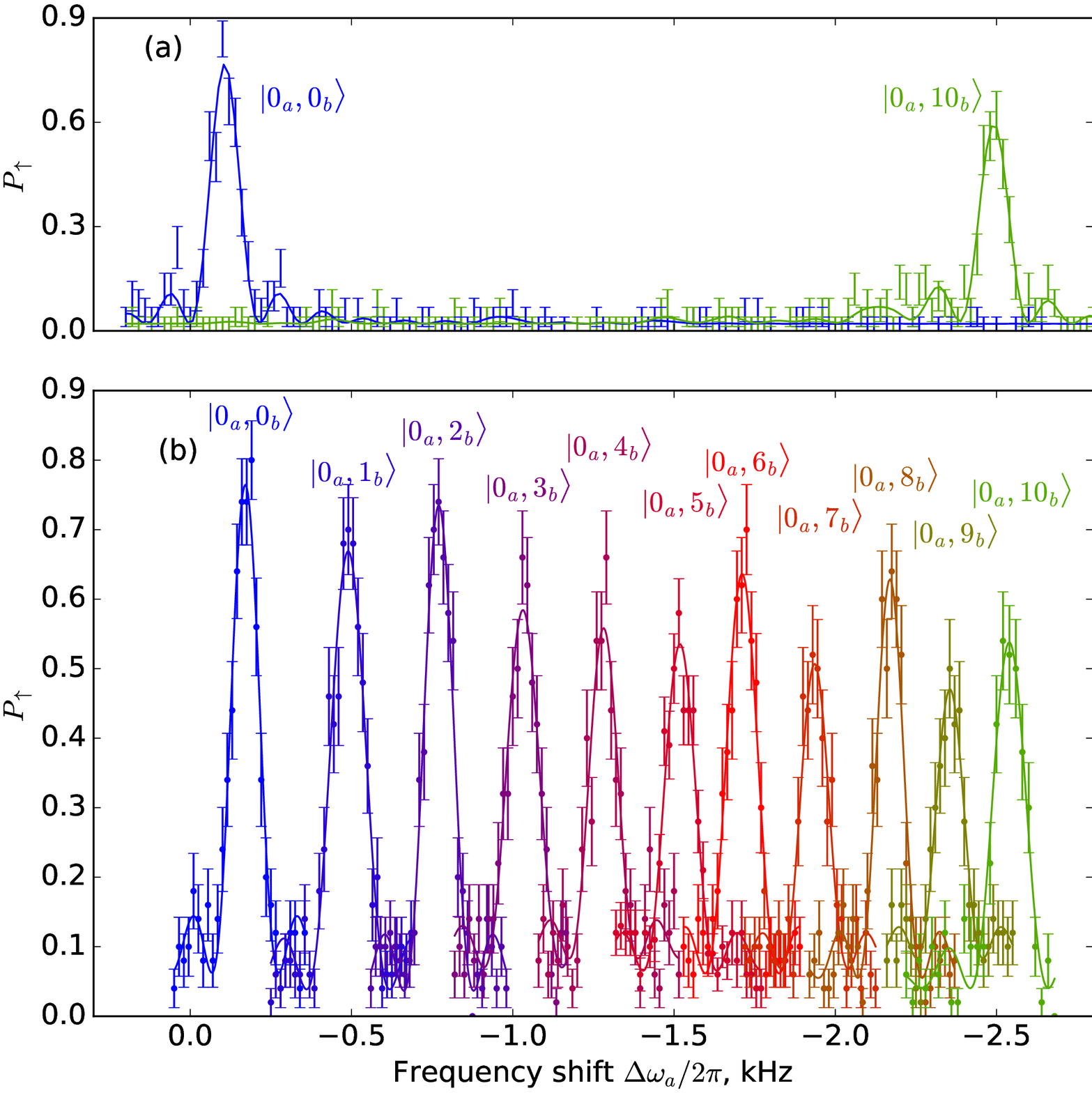}
\includegraphics[width=0.98 \columnwidth]{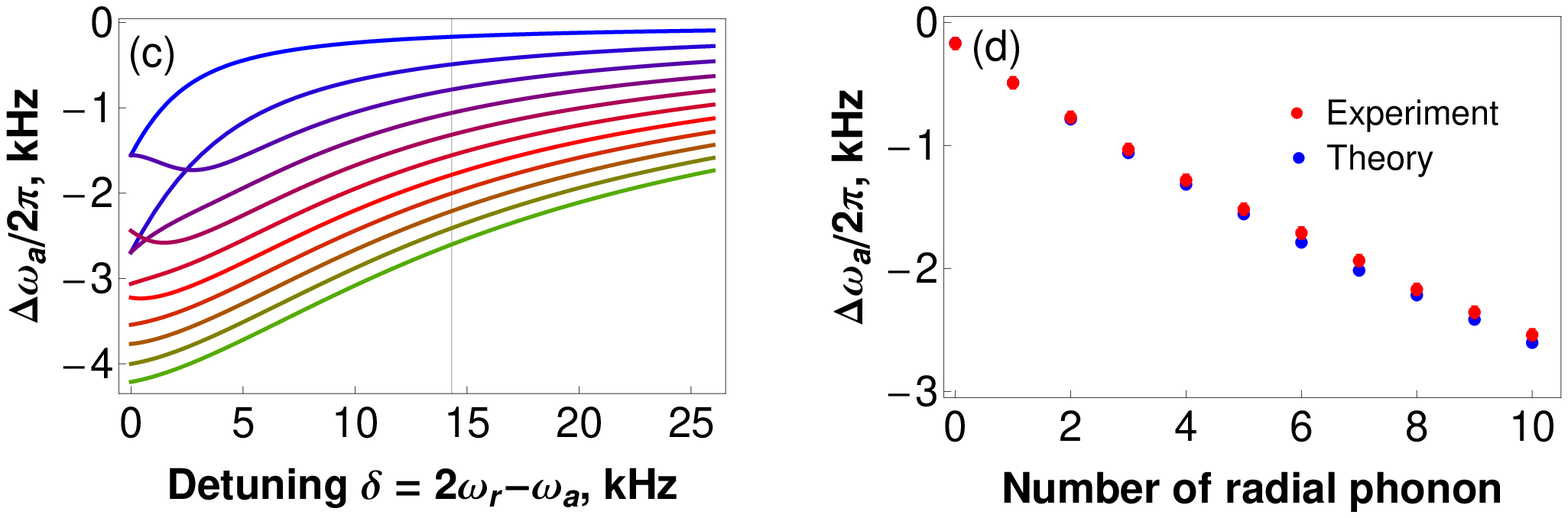}
\caption{\label{fig:fock_peaks}
(a) Full frequency scan of the axial blue sideband when the system is prepared 
in Fock states $|0_a, 0_b\rangle$ (blue) and $|0_a, 10_b\rangle$ (green).
For the state $|0_a, 10_b\rangle$, the reduced height of the main peak and residual 
population detected in the states with $n_b<10$ are due to imperfections of the state preparation
that, according to the fit, populates the state $|0_a, 10_b\rangle$ with probability 
$p_{10} = 0.80(2)$, and 
leaves significant populations $p_{9} = 0.06(2)$, and  $p_{8} = 0.06(2)$ in the states
$|0_a, 9_b\rangle$ and $|0_a, 8_b\rangle$. Preparation of state $|0_a, 0_b\rangle$
populates other Fock states with the probability $<0.04$. 
(b) Measured position of the axial motional sideband for
the Fock states $|0_b\rangle$ to $|10_b\rangle$ prepared in the radial mode. Axial mode is 
initially prepared in the vacuum state $|0_a\rangle$.
Only part of the data around the largest peak is shown for each state. 
(c) Calculated shift of the axial sideband frequency as a function of detuning $\delta$ 
for different states ($|0_a, 0_b\rangle$ to $|0_a, 10_b\rangle$).
(d) Calculated and measured shift of the axial sideband frequency as a function of 
the number of phonons added to the radial mode for detuning $\delta / 2 \pi = 14.3$\,kHz. 
}
\end{figure}

When the axial and radial modes are far detuned from each other, i.e., $\delta \gg \xi$, 
the coupling manifests itself as frequency shifts of the modes. 
In this limit, the  shift of the axial mode $\delta E =-2 (2n_b+1) \xi^2/\delta$ is proportional to the number of phonons $n_b$ in the radial mode.
In our experiment this shift deviates from a linear relation due to finite value of $\delta$.  
However, exact diagonalization of the Hamiltonian Eq.(\ref{eq:hamiltonian}) confirms that the 
frequency shift is still a monotonous function of the radial phonon number $n_b$, as shown in Fig.\ref{fig:fock_peaks}(c).
In the following, we employ this effect to implement the projective measurement of the radial motional state in the Fock basis. 

To  prepare the Fock state $|n_b\rangle$ in the radial mode we start from the ground state of motion and apply a series of $\pi$-pulses on 
$|{\downarrow}\rangle |n_b\rangle \rightarrow 
 |{\uparrow}\rangle |n_b+1\rangle  $ and 
$|{\uparrow}\rangle |n_b\rangle \rightarrow 
 |{\downarrow}\rangle |n_b+1\rangle  $ sideband transitions followed by optical pumping to reinitialize the ion back to the $|{\downarrow}\rangle$ state~\cite{ourpaper_parametricoscillator,1996PRLNonclassicalstate}. 
We set the detuning $\delta / 2 \pi = 14.3$\,kHz and Raman beams power such that the time taken to apply a $\pi$ pulse 
for the first order axial blue sideband transition is $T_{\pi}=8$\,ms. This ensures that the transform limited width of the axial sideband peak is smaller than the axial sideband shift 
due to the presence of a single phonon in the radial mode.
The axial blue sidebands for different Fock states with $n_b=0$ to $n_b=10$ in the radial mode
are presented in Fig.\ref{fig:fock_peaks}(b). 

The central frequency $\omega_{n}$ of the axial sideband is determined by fitting the peaks with
$p_n(\omega) = g + \eta f_n(\omega)$, where 
$f_n (\omega) = [(\Omega / \Omega_n) \mathrm{sin}(\pi \Omega_n / 2 \Omega)]^2$, 
$\Omega = \pi / T_{\pi}$ is the Rabi frequency of the transition, 
$\Omega_n = \sqrt{\Omega^2 + \Delta_n^2}$, $\Delta_n = \omega - \omega_n$ 
is the detuning of the driving frequency from the resonance, 
$\eta$ is the phonon detection efficiency, $g$ describes the state detection background contribution and $n$ is the number of phonons in the radial mode. The observed frequency shifts are on the order of 300 Hz / phonon and are in 
good agreement with the theoretical predictions calculated by diagonalization  of the Hamiltonian Eq.(\ref{eq:hamiltonian}),
as shown in Fig.\ref{fig:fock_peaks}(c-d). The peaks are clearly separated from each other, 
which makes it possible to efficiently detect the phonon number distribution for 
the radial mode. 

\begin{figure*}[tbp!]
\centering
\includegraphics[width=2.0 \columnwidth]{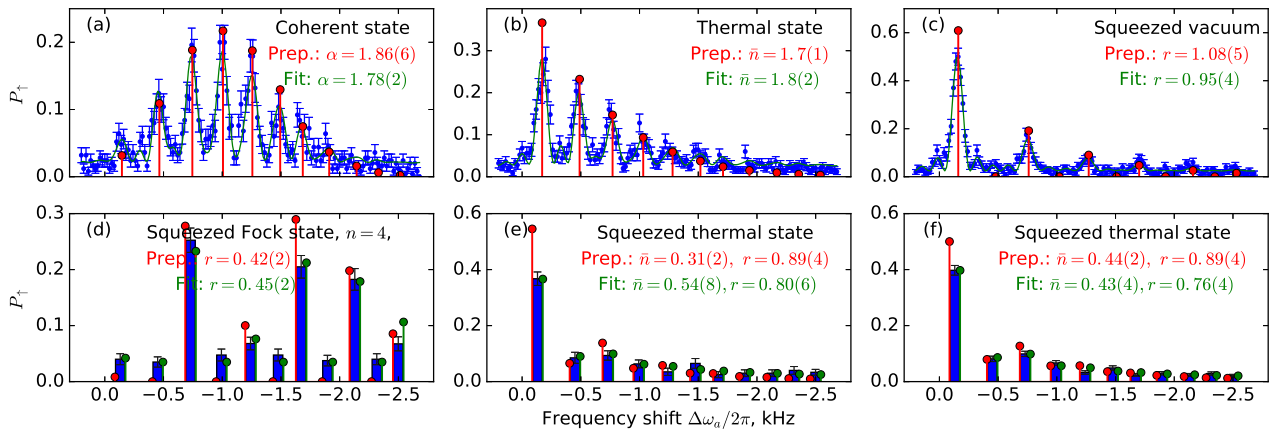}
\caption{\label{fig:states_freq_scan}
Top row: frequency scans of axial blue sideband for the coherent state (a), 
thermal state (b)
and squeezed vacuum state (c). 
Resolved peaks are clearly visible. Solid red bars indicate the expected probability $p_n$ to have $n$ phonons for a given quantum state. Here state parameters were obtained from an independent calibration. Green lines shows the fit of the experimental data for the corresponding state.
Bottom row: reconstructions of phonon number distribution for the squeezed Fock state (d), 
squeezed thermal states (e) 
and (f) are obtained by the projective measurement as described in the text. 
The state parameters extracted from the fit of the experimental data (green) 
are in reasonable 
agreement with the independent calibration of the preparation procedure (red). 
The errors correspond to $1\sigma$ statistical uncertainties.}
\end{figure*}

Such measurements for the coherent, thermal and squeezed states are shown in  Fig.\ref{fig:states_freq_scan}(a-c). 
To prepare the coherent state $|\alpha\rangle = e^{-|\alpha|^2 / 2} \sum_{n} \alpha^{n} |n\rangle / \sqrt{n!}$, 
we first initialize the bright ion in the $|^2S_{1/2}, F=0, m_F=1\rangle $ state and then apply an optical dipole force
modulated at the mode frequency $\omega_b$~\cite{ourpaper_parametricoscillator,microwave}. 
The thermal state is prepared by applying a few (18 in the case shown in Fig.\ref{fig:states_freq_scan}(b)) coherent displacement pulses with duration 
 100 $\mu s$, each with a random phase. The resulting random walk in phase space leads to a thermal state 
$\rho = \sum_n |n\rangle \langle n |\, \bar n^{n}/ (1 + \bar n)^{1+n}$ \cite{Loudon_book}, where $\bar{n}$ is the mean phonon number.
The squeezed vacuum state $\hat{S}(r)|0_b\rangle$, where 
$\hat{S}(r) = e^{ (r b^2 + r^* b^{\dagger 2} ) /2 } $ is a squeezing operator and
$r$ is a squeezing parameter, is prepared by modulating the optical dipole force 
at twice the mode frequency $2 \omega_b$~\cite{1996PRLNonclassicalstate}. 

The height of each peak in the measured axial sideband spectra is proportional to the corresponding phonon population. To determine the state 
parameters, the data shown in Fig.~\ref{fig:states_freq_scan} are fitted with $p(\omega) = g + \eta \sum_{n=0}^{10} p_n f_n(\omega)$, while allowing parameters $\alpha$, $\bar{n}$ or $r$ to vary. Here $p_n$ is the expected phonon number distribution for the corresponding state.
The results are in good agreement with independent calibrations of the preparation procedures. This fitting procedure also determines the efficiency of the single shot measurement $\eta=0.70(2)$, i.e. probability to detect the ion in the state $|{\uparrow}\rangle$ given the Fock state $|n\rangle$ was prepared and measured. This efficiency is limited mostly by fast fluctuations of the Raman laser power and pointing stability of the laser beams. 

Since the positions of the axial blue sideband $\omega_n$ as a function of the radial phonon number $n$ 
are known (Fig. 2(d)), we can perform the projective measurement of a radial mode state and determine whether it is in the Fock state with $n$ phonons in a single shot. 
This is achieved by applying a $\pi$ pulse at frequency $\omega_n$ followed 
by fluorescence detection of the ion internal state.  If no light is detected, the initial state of the radial mode $|\Psi\rangle$ is projected to a state $P_n |\Psi\rangle$ by a projector $P_n = 1 -|n\rangle \langle n|$~\cite{ourpaper_parametricoscillator}. If fluorescence is detected, the mode is determined to be in Fock state $|n\rangle$. Our experiment deviates from ideal projective measurement because, in the latter case, fluorescence of the ion destroys its motional state. However, since the outcome of the measurement is known, this state can, in principle, be prepared again~\cite{monroe_qnd_2011}. 

The procedure described above is performed on squeezed thermal $\hat{S}(r) \rho \hat{S}^{\dagger}(r)$ and squeezed Fock $\hat{S}(r) |n\rangle \langle n| \hat{S}^{\dagger}(r)$ states. These states are prepared by applying the squeezing operation to 
the thermal and Fock states, respectively. The results are shown in Fig.\ref{fig:states_freq_scan}(d-f).
The parameters $\bar{n}$ and $r$ obtained from the fit of the experimental data agree well with the expected values.

Some discrepancies between the measurements in Fig.\ref{fig:fock_peaks}, Fig.\ref{fig:states_freq_scan} and the expected values can be attributed to the infidelities of motional state preparation, imperfect driving of the axial sideband transition and detection of the ion internal state, decoherence of ion motion during the measurement~\cite{ourpaper_parametricoscillator}, as well as residual mixing of ``bare'' eigenstates due to the coupling $H_c$ for the given two-mode detuning $\delta$.

The results demonstrated in this letter may be of use in quantum information processing with  
continuous variables~\cite{jiang_2015,holland_cross_kerr_2015} and qubits~\cite{yamamoto_1995,duan_2000}, preparation of nonclassical states of 
motion~\cite{paternostro_cat_2003} via  projective measurements~\cite{Schoelkopf_2014}, serve as a fast alternative method to extract 
the phonon number distribution~\cite{1996PRLNonclassicalstate,Leibfried_tomo_1996}, 
fully reconstruct the quantum state~\cite{Haroche_2008,Schoelkopf_2014}
and shed light on the  cross-Kerr coupling problem in quantum optics~\cite{shapiro_kerr_2006,shapiro_kerr_2007,gea_banacloche_kerr_2010,fan_kerr_breakdown_2013}.

\begin{acknowledgments}
We thank Jiang Liang and Atac Imamoglu for helpful discussions and Brenda Chng for help with the manuscript preparation. 
This research is supported by the National Research Foundation, Prime Minister’s Office, Singapore and the Ministry of Education, Singapore under the Research Centers of Excellence program and Education Academic Research Fund Tier 2 (Grant No. MOE2016-T2-1-141).
\end{acknowledgments}


%

\end{document}